\documentclass[10pt, final]{IEEEtran}
\usepackage{epsfig,cite,stfloats,bm,amssymb,amsmath,color,subfigure,graphics,extpfeil}

\long\def\symbolfootnote[#1]#2{\begingroup
\def\thefootnote{\fnsymbol{footnote}}
\footnote[#1]{#2}\endgroup}
\psfull

\begin{document}

% Title.
% ------
\title{Flexible Pilot Contamination Mitigation with Doppler PSD Alignment}

\author{\normalsize{Xiliang Luo and Xiaoyu Zhang}

\thanks{This work was supported through the startup fund from ShanghaiTech
University under the grant no. F-0203-14-008.}

\thanks{Xiliang Luo and Xiaoyu Zhang are with the School of Information Science and Technology, ShanghaiTech University, 319 Yueyang Road, Shanghai, 200031, China. Tel/fax:
+86-21-54205213/54203396, Email: {\tt \{luoxl, zhangxy\}{\rm\char64}shanghaitech.edu.cn}}
}

% make the title area
\maketitle

\markboth{{\scriptsize IEEE Signal Processing Letters (AQ)}}{}
\renewcommand{\thepage}{}

%\hspace*{4.0cm}
%\parbox{\textwidth}{
%{
%\begin{tabular}{rl}
%{\small\bf Suggested EDICS:}& COM-MAMI \\
%{\small\bf Suggested Associated Editorial Areas:} & {\tt Massive MIMO}\\
%{} & {\tt Channel estimation}\\
%{\small\bf Special Issue:}& Energy-Efficient Techniques for $5$G Wireless\\
%{} & Communication Systems\\
%{\small\bf ID Number:} & {\tt TW-Jul-15-0930}\\
%{\small\bf Original Submission:} & {\tt June 8, 2016} \\
%{\small\bf 1st Revision (AQ):} & {\tt \today}\\
%{\small\bf 2nd Revision:} & {\tt January 27, 2016}\\
%{\small\bf Accepted:} & {\tt March 8, 2016}\\
%{\small\bf Editor:} & {\tt Dr. Jun Zhang}\\
%{\small\bf Email:} & {\tt eejzhang@ust.hk}\\
%\end{tabular}}}

\newpage
\pagenumbering{arabic}\setcounter{page}{1} \markboth{{}}{}

\begin{abstract}
Pilot contamination in the uplink (UL) can severely degrade the channel estimation
quality at the base station (BS) in a massive multi-input multi-output (MIMO)
system. Thus, it is critical to explore all possible avenues to enable more
orthogonal resources for the users to transmit non-interfering UL pilots. In
conventional designs, pilot orthogonality typically assumes constant channel
gains over time, which limits the amount of orthogonal resources in the case
of time-selective channels. To circumvent this constraint, in this paper, we show
how to enable orthogonal multiplexing of pilots in the case of Doppler fading by
aligning the power spectrum densities (PSD) of different users. From the derived
PSD aligning rules, we can see multiple users can be sounded simultaneously
without creating/suffering pilot contamination even when these users are
experiencing time-varying channels. Furthermore, we provide analytical formulas
characterizing the channel estimation mean square error (MSE) performance.
Computer simulations further confirm us the PSD alignment can serve as one
important decontamination mechanism for the UL pilots in massive MIMO.
\end{abstract}


\section{Introduction}
% Massive MIMO overview
By deploying a large number of antennas at the base station (BS),
massive multiple-input multiple-output (MIMO) will be able to bring
significant spectral efficiency gains. It has been regarded as one
of the key enabling technologies for the next generation wireless
communications \cite{marzetta10twc, resek13spmag,larsson14commag,lu14jstsp}.
To ensure best channel estimation quality, it is desirable to allocate
orthogonal uplink (UL) pilot sequences to different users so that the pilot
transmissions do not interfere with each other. But within a limited
time period and a limited bandwidth, there are only a limited number of
orthogonal pilot sequences. As the number of users becomes large,
non-orthogonal pilot sequences need to be re-used by the users
served by different BSs, which leads to the so-called pilot contamination
\cite{marzetta10twc,lu14jstsp}. Pilot contamination is one severe
limiting factor in multi-cell massive MIMO systems.

Various approaches have been proposed to alleviate the pilot contamination issue
in massive MIMO. Recent works include
\cite{marzetta13jsac,jin2016tvt, luo2016globecom,yin2013jsac,yin2014jstsp,xia2015twc,swindlehurst2016tsp,yin2016tsp,hu16twc,muller14jstsp,Upadhya2016}.
By staggering the UL transmission timeline of different cells, the time-shifted
pilots were proposed in \cite{marzetta13jsac} to mitigate the pilot contamination
and were further analyzed in \cite{jin2016tvt}.
In \cite{yin2013jsac,yin2014jstsp,xia2015twc,yin2016tsp}, pilot decontamination was achieved
by utilizing the fact that users with non-overlapping angles of arrival (AoA) enjoy
asymptotic orthogonal covariance matrices.
Phase shift pilots were exploited for channel acquisition in massive MIMO systems
employing orthogonal frequency division multiplexing (OFDM)
\cite{swindlehurst2016tsp} and to mitigate the pilot contamination by aligning the
channel power distributions in the delay-angle domain \cite{luo2016globecom}.
Blind methods were proposed in
\cite{muller14jstsp,hu16twc} and the pilot contamination effect was shown to
diminish as the data length grew. In \cite{Upadhya2016}, to multiplex more
orthogonal pilots without losing the dimensionality for data transmission,
superimposed pilots \cite{Tugnait2007} were proposed for massive MIMO where pilots
were sent together with data. However, constant channels were assumed in
\cite{Upadhya2016} to ensure enough processing gain during channel estimation to
combat the data interference. In fact, constant channel gains over time are
typically assumed in conventional pilot (de)contamination studies
\cite{lu14jstsp,jose2011,marzetta10twc}.

In this paper, we explore a new avenue to enable orthogonality among users' UL
pilots even when the users' channels are time-varying. We demonstrate that
orthogonal multiplexing of pilots in the case of Doppler fading can be achieved by
aligning the Doppler power spectrum densities (PSD) of different users judiciously.
Furthermore, the proposed PSD alignment enables flexible mitigation of time-varying
inter-cell pilot contamination. Meanwhile, we are also able to characterize the
channel estimation mean square error (MSE) performance analytically with the Doppler
PSDs.

% Structure
%This paper is organized as follows: Section \ref{SecSysModel} describes the system
%model and provides the general orthogonality conditions. Section
%\ref{secDopplerDesigns} details the orthogonal pilot designs in the case of Doppler
%fading channels. Channel estimation performance is then characterized in
%\ref{secOrthPerf}. Corroborating simulation results are provided in Section
%\ref{secPerformance} and Section \ref{conclusion} concludes the paper.

{{\it Notations:} ${\sf Diag}\{\cdots\}$ denotes the diagonal matrix with diagonal
elements defined inside the curly brackets. ${A}(i,j)$ refers to the $(i,j)$th
entry of matrix $\bm A$ and $a(i)$ stands for the $i$-th entry of the vector ${\bm a}$. $\bm I$ denotes the identity matrix. ${\sf E}[\cdot]$, ${\tt Tr}(\cdot)$, $(\cdot)^{\dagger}$, $(\cdot)^{T}$, and $(\cdot)^*$ represent
expectation, matrix trace, Hermitian operation, transpose, and conjugate operation respectively.}

\section{System Model and Orthogonality Conditions}\label{SecSysModel}
In a typical massive MIMO system, each BS is equipped with $M$ antennas
and $K$ single-antenna users\footnote{These $K$ users include all the users
served by all the BSs.}
send UL pilots simultaneously. Assuming a narrow-band
channel, e.g. a particular subcarrier in the case of OFDM transmission, we can have
the following system model in the UL over $P$ successive pilot slots at one
particular BS:
\begin{equation}\label{cha mod}
{\bm y}_{m}=\sum_{k=1}^{K}\sqrt{\rho_k}{\bm X}_{k}{\bm h}_{k,m}+{\bm w}_{m},
\end{equation}
where ${\bm y}_{m}\in {\mathbb C}^P$ stands for the received signal vector at the
$m$th antenna of the BS over $P$ pilot slots, $\rho_k$ denotes the transmitted power
from user-$k$, ${\bm X}_k={\sf Diag}\{x_k(0),...,x_k(P-1)\}$ contains the
transmitted pilots within the successive $P$ pilot slots from the $k$th user, ${\bm h}_{k,m}=[h_{k,m}(0),...,h_{k,m}(P-1)]^T$ represents the channel vector from
user-$k$ to the $m$th antenna at the BS over the $P$ time slots, and
${\bm w}_{m}\in {\mathbb C}^P$ stands for the additive receiver noise. Here, we make the following assumptions:
\begin{itemize}
  \item{\bf AS1}: The channel samples: $\{h_{k,m}(n)\}$ form a unit-power
  Gaussian stationary process with the autocorrelation function defined
  as\footnote{Since PSD is the Fourier
  transform of $r_k(v)$, we are assuming the same PSD for different receive antennas.}: $r_k(v):={\sf E}[h_{k,m}(l)h_{k,m}(l+v)^*]$. Let ${\bm R}_k:={\sf E}[{\bm h}_{k,m}{\bm h}_{k,m}^{\dag}]$ denote the covariance of the channel vector ${\bm h}_{k,m}$. We have $R_{k}(l,l')=r_k(l'-l)$. Meanwhile, the channels from different users are assumed independent to each other, i.e. ${\sf E}[{\bm h}_{k,m}{\bm h}_{g,m}^{\dag}]={\bm R}_k\delta(k-g)$;
  \item{\bf AS2}: The receiver noise ${\bm w}_m$ is zero mean and circularly
  symmetric Gaussian with covariance matrix:
  ${\sf E}[{\bm w}_m{\bm w}_m^{\dag}] = \sigma^2{\bm I}_{P}$;
  \item{\bf AS3}: The pilot sequence ${\bm X}_k$ enjoys constant unit modulus,
  i.e. ${\bm X}_k{\bm X}_k^{\dag}={\bm I}$.
\end{itemize}
According to (\ref{cha mod}), we can readily obtain the MMSE estimate of user-$k$'s
channel as follows:
\begin{equation}\label{cha est}
\begin{aligned}
  {\hat{\bm h}}_{k,m}&={\sf E}[{\bm h}_{k,m}{\bm y}_{m}^{\dag}]\left({\sf E}[{\bm y}_{m}{\bm y}_{m}^{\dag}]\right)^{-1}{\bm y}_{m}\\
  &=\sqrt{\rho_k}{\bm R}_k{\bm X}_k^{\dag}\cdot\\
  &\left(\sigma^2{\bm I}+\rho_k{\bm X}_k{\bm R}_k{\bm X}_k^{\dag}+
  \sum_{g=1,g\ne k}^{K}
  \rho_g{\bm X}_g{\bm R}_g{\bm X}_g^{\dag}\right)^{-1}{\bm y}_{m}.
\end{aligned}
\end{equation}
The covariance of the channel estimation error:
${\bm \epsilon}:={\bm h}_{k,m}-{\hat{\bm h}}_{k,m}$ can be expressed as follows:
\begin{equation}\label{err int}
  \begin{aligned}
  &{\sf E}[{\bm \epsilon}{\bm \epsilon}^{\dag}]=
  {\bm R}_k-\rho_k{\bm R}_k{\bm X}_k^{\dag}\cdot\\
  &\left(\sigma^2{\bm I}+\rho_k{\bm X}_k{\bm R}_k{\bm X}_k^{\dag}+
  \sum_{g=1,g\ne k}^{K} \rho_g{\bm X}_g{\bm R}_g{\bm X}_g^{\dag}\right)^{-1}{\bm X}_k{\bm R}_k\\
  &=
  {\bm R}_k-\rho_k{\bm R}_k\left(\sigma^2{\bm I}+\rho_k{\bm R}_k+{\bm \Delta}\right)^{-1}{\bm R}_k,
  \end{aligned}
\end{equation}
where
$${\bm \Delta}:=\sum_{g=1,g\ne k}^{K}\rho_g
{\bm X}_k^{\dag}{\bm X}_g{\bm R}_g{\bm X}_g^{\dag}{\bm X}_k$$
contains the interference from other users' pilots. In the absence of those
interference terms, the corresponding channel estimation MSE is
\begin{equation}\label{err ni}
  {\sf E}[{\bm \epsilon}_0{\bm \epsilon}_0^{\dag}]={\bm R}_k-\rho_k{\bm R}_k\left(\sigma^2{\bm I}+\rho_k{\bm R}_k\right)^{-1}{\bm R}_k.
\end{equation}
In order to obtain the interference-free MSE performance as shown in (\ref{err ni}),
we can establish the following proposition:\\
\noindent{\bf Proposition 1:}{\it Under AS1$\sim$3, the interference-free channel estimation performance in (\ref{err ni}) is achieved when the pilot sequences of unit modulus
satisfy the following orthogonality conditions:
\begin{equation}\label{ortho}
  {\bm R}_k{\bm P}_{k,g}{\bm R}_g{\bm P}_{k,g}^{\dag}={\bm 0}, \forall g\neq k,
\end{equation}
where ${\bm P}_{k,g}={\bm X}_k^{\dagger}{\bm X}_g$. }

Conventional orthogonal designs of the pilot sequences assume that the channel remains
constant over the $P$ time slots containing UL pilots, where we have
${\bm R}_k={\bm 1}\cdot{\bm 1}^{\dag}$ and ${\bm 1}:=[1,...,1]^T$. Thus, the
orthogonality condition in (\ref{ortho}) naturally becomes: $\forall g\neq k$,
\begin{equation}\label{ortho2}
 \begin{aligned}
 &{\bm 1}\cdot{\bm 1}^{\dag} {\bm P}_{k,g}{\bm 1}\cdot{\bm 1}^{\dag}{\bm P}_{k,g}^{\dag} = ({\bm 1}^{\dag} {\bm P}_{k,g}{\bm 1}){\bm 1}\cdot{\bm 1}^{\dag}{\bm P}_{k,g}^{\dag}\\
 &={\tt Tr}({\bm P}_{k,g}){\bm 1}\cdot{\bm 1}^{\dag}{\bm P}_{k,g}^{\dag}={\bm 0}\\
 &\Leftrightarrow {\tt Tr}({\bm P}_{k,g})=0.
 \end{aligned}
\end{equation}
It can be easily seen that the above result simply informs us that the pilot
sequences should be designed such that the inner product between each pair is zero.
Although enjoying simplicity, the underlying assumption of a constant channel
across the $P$ time slots severely limits the multiplexing capability of the
conventional orthogonal designs in the case of Doppler, when the channel can
be regarded constant only within a small portion of the channel coherence time \cite{DigitalComBook}.
In the following sections, we will address the orthogonal designs in the presence
of Doppler shifts.

\section{Orthogonal Designs via PSD Aligning}\label{secDopplerDesigns}
As $P$ gets large, we can approximate ${\bm R}_k$ by a circulant matrix ${\bm C}_k$,
whose first column is defined as follows \cite{adhikary13tit, Gray2006}:
\begin{equation}
  C_k(:,1)=
  \left[\begin{array}{c}
  r_k(0)\\
  r_k(-1)+r_k(P-1)\\
  r_k(-2)+r_k(P-2)\\
  \vdots\\
  r_{k}(-P+1)+r_k(1)\end{array}\right].
\end{equation}
The eigenvalue decomposition (EVD) of ${\bm C}_k$ can be expressed as
${\bm C}_k={\bm F}^{\dag}{\bm \Lambda}_k{\bm F}$, where ${\bm F}$ is the unitary
$P$-point FFT matrix and ${\bm \Lambda}_k$ contains the eigenvalues.
To achieve orthogonality between the UL pilots from different users in the presence
of Doppler shifts, the unit modulus pilot sequences need to satisfy the condition
specified in (\ref{ortho}). By approximating ${\bm R}_k$ with ${\bm C}_k$, we can
rewrite the condition in (\ref{ortho}) as
\begin{equation}\label{ort}
\begin{aligned}
  {\bm R}_k{\bm P}_{k,g}{\bm R}_g{\bm P}_{k,g}^{\dag}&\approx
  {\bm F}^{\dag}{\bm \Lambda}_k{\bm F}{\bm P}_{k,g}{\bm F}^{\dag}{\bm \Lambda}_g{\bm F}{\bm P}_{k,g}^{\dag}\\
  &={\bm F}^{\dag}{\bm \Lambda}_k{\bm \Theta}_{k,g}{\bm \Lambda}_g{\bm \Theta}_{k,g}^{\dag}{\bm F}={\bm 0},
\end{aligned}
\end{equation}
where ${\bm \Theta}_{k,g}:={\bm F}{\bm P}_{k,g}{\bm F}^{\dag}$. According to
(\ref{ort}), the following requirement on the pilot sequences can be established:\\
\noindent{\bf Proposition 2}:{\it Under AS1$\sim$3, as the length of the channel
observations: $P$ becomes large, the UL pilots between user-$k$ and
user-$g$ at each receive antenna at the BS become orthogonal when the following
condition is met:
\begin{equation}\label{ort con}
  {\bm \Lambda}_k{\bm \Theta}_{k,g}{\bm \Lambda}_g{\bm \Theta}_{k,g}^{\dag}={\bm 0}.
\end{equation}}

Motivated by the structure of ${\bm \Theta}_{k,g}={\bm F}{\bm P}_{k,g}
{\bm F}^{\dag}$ in (\ref{ort con}), we consider the following FFT pilot sequences:
\begin{equation}
{\bm X}_k = {\sf Diag}\left\{1,e^{j\frac{2\pi\tau_{k}}{P}},...,e^{j\frac{2\pi\tau_{k}(P-1)}{P}}
\right\}\cdot{\bm S}_0, \label{CyclicShift}
\end{equation}
where $\tau_{k}$ is the amount of cyclic time shifts and ${\bm S}_0$ is the base
unshifted sequence with constant modulus. Note the above designs have been exploited
in the LTE UL \cite{LTEBook}. Then the matrix ${\bm \Theta}_{k,g}$ becomes unitary and circulant with the first column vector taking the following form:
\begin{eqnarray}
{\bm \Theta}_{k,g}(:,1)^T=
[\underbrace{0,...,0}_{\Delta\tau},1,\underbrace{0,...,0}_{P-\Delta\tau-1}],\label{FirstColumn}
\end{eqnarray}
where $\Delta\tau:=\tau_{g}-\tau_{k}$ refers to the amount of relative cyclic
shifts between the user-$g$ and user-$k$. Accordingly, the diagonal matrix
$\tilde{{\bm\Lambda}}_g:={\bm\Theta}_{k,g}{\bm\Lambda}_g{\bm\Theta}_{k,g}^{\dag}$
is obtained by cyclicly shifting the diagonals\footnote{Positive value means cyclic
shifts towards the bottom right.} of ${\bm\Lambda}_g$ by
an amount of $\Delta\tau=\tau_g-\tau_k$. From Proposition 2, we can further have
the following corollary:\\
\noindent{\bf Corollary 2.1}: {\it Under AS1$\sim$3, as the length of the channel
observations: $P$ goes large, with the FFT pilot sequences in (\ref{CyclicShift}),
the received UL pilots from user-$k$ and user-$g$ at the BS become orthogonal when
the following condition is met:
\begin{equation}
{\bm \Lambda}_k {\mathbb S}_{\tau_g-\tau_k}\{{{\bm \Lambda}}_g\}={\bm 0},
\label{shiftCond}
\end{equation}
where ${\mathbb S}_{\tau}\{\cdot\}$ stands for the operation of shifting
the diagonal elements of the argument cyclicly by the amount of $\tau$.}

\begin{figure}[t]
\centering
\epsfig{file=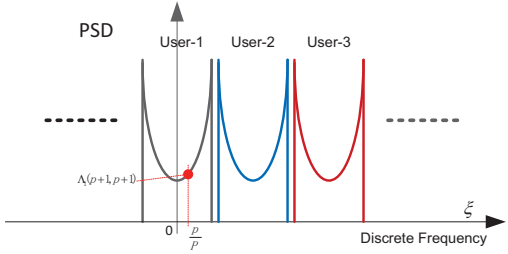,width=0.4\textwidth}
\caption{Pilot orthogonality via PSD alignment.}
\label{FigPSDAlign}
\vspace{-0.5cm}
\end{figure}

The PSD of the sampled channel process $\{h_{k,m}(n)\}$ can be obtained by computing
the discrete-time Fourier transform (DTFT) of the autocorrelation sequence $r_k(v)$,
i.e. $S_k(\xi)=\sum_{v=-\infty}^{\infty}r_k(v)e^{-j2\pi\xi v}$.
As $P$ becomes large, the $P$ eigenvalues of ${\bm C}_k$ can be approximated as the
uniformly spaced samples of $S_k(\xi)$: $\{S_k([p/P]_{(-1/2,1/2]}): p=0,1,...,P-1\}$, where $[\cdot]_{(-1/2,1/2]}$ refers to the modulo operation such
that the result lies in the interval $(-1/2,1/2]$. Let $f_{D,k}$ denote the
maximum Doppler frequency of user-$k$. When the channel process $\{h_{k,m}(n)\}$ is
obtained by sampling the underlying continuous-time fading channel at a sampling
frequency of $f_s$, we know $S_k(\xi)$ is band-limited within $[-F_k, F_k]$ with
$F_k:=f_{D,k}/f_s$ representing the maximum discrete frequency of the PSD
$S_k(\xi)$. In a nutshell, Corollary 2.1 informs us that, in order to ensure orthogonal pilots between two users, the amount of the relative cyclic shifts
between the two users should be judiciously chosen such that the supports of their
shifted PSDs are non-overlapping (see also Fig. \ref{FigPSDAlign}).
When all the users share the same maximum discrete Doppler shift $F$, we can see up
to $\frac{1}{2F}$ users can transmit orthogonal pilots simultaneously and the
relative cyclic shift values among users are $\left\{k\cdot 2FP, k=0,1,...,\frac{1}{2F}-1\right\}$ correspondingly.

In addition to being able to support multiple orthogonal pilots in time-selective
channels, the cyclic shifts in (\ref{CyclicShift}) can also be flexibly chosen to
dodge the pilot contamination from other cells. For example, when we know there
exists strong interference over the frequency supports of user-$1$ and user-$2$ in
Fig. \ref{FigPSDAlign}, we will choose the cyclic shifts such that the shifted PSDs
of the desired users are non-overlapping with the interference. In Section
\ref{secPerformance}, we will simulate this situation and verify the effectiveness
of our designs.

\section{Orthogonal Performance with Doppler}\label{secOrthPerf}

In the case of multiple users, we can get the following result characterizing the
channel estimation MSE performance (see detailed proof in Appendix A):\\
\noindent{\bf Proposition 3}: {\it When the pilot sequences are designed as in
(\ref{CyclicShift}), as $P$ goes large, under AS1$\sim$3, the MSE of each
element of ${\bm h}_{k,m}$ can be approximated as follows:
\begin{equation}\label{mse gen}
  \lim_{P\rightarrow\infty}\hspace{-0.1cm}{\sf MSE}_k =
  1-\hspace{-0.2cm}\int_{-\frac{1}{2}}^{\frac{1}{2}}\frac{S_k^2(\xi)\rho_k}{S_k(\xi)\rho_k+\sum_{g=1,g\ne k}^{K}\tilde{S}_{g}(\xi)\rho_g+\sigma^2} d\xi,
\end{equation}
where $S_k(\xi)$ denotes the PSD of user-$k$, and $\tilde{S}_g(\xi):=S_{g}(\xi-(\tau_g-\tau_k)/{P})$.}

Assuming the Jakes's fading model \cite{JakesModel} (a.k.a. Clarke's model), the
autocorrelation function: $r_k(v)$ can be expressed as
$r_k(v)=J_0(2\pi F_k v)$, where $f_{D,k}$ denotes the maximum Doppler frequency of
user-$k$, $f_s$ represents the channel sampling frequency, $F_k:=f_{D,k}/f_s$ is the
normalized Doppler frequency, and $J_0(\cdot)$ is the zeroth-order Bessel function
of the first kind. When $F_k\le\frac{1}{2}$, the PSD of the discrete channel
process: $\{h_{k,m}(n)\}$ can be expressed as follows:
\begin{equation}\label{psd_c}
  S_k(\xi)=\left\{\begin{array}{cc}\frac{1}{\pi}\cdot\frac{1}{\sqrt{F_k^2-\xi^2}}, &\quad \xi\in\left[-F_k,F_k\right]\\
  0,&\quad \xi\in[-\frac{1}{2},-F_k]\cup[F_k,\frac{1}{2}].\end{array}\right.
\end{equation}
By combining (\ref{psd_c}) and (\ref{mse gen}), as the orthogonality conditions in
(\ref{ortho}) or (\ref{shiftCond}) are satisfied, the channel estimation MSE in
(\ref{mse gen}) can be approximated as follows:
\begin{equation}\label{mse c}
  \lim_{P\to\infty}{\sf MSE}_k=\left\{\begin{array}{cc} 1-\frac{4}{\pi\sqrt{1-\alpha^2}}\arctan\sqrt{\frac{1-\alpha}{1+\alpha}},&\alpha<1\\
  1-\frac{2}{\pi\sqrt{\alpha^2-1}}\ln\left|\frac{\alpha-1+\sqrt{\alpha^2-1}}{\alpha-1-\sqrt{\alpha^2-1}}\right|,&\alpha>1\\
  1-\frac{2}{\pi},&\alpha=1
  \end{array}\right.
\end{equation}
where $\alpha:=\pi F_k\sigma^2/\rho_k$.

Let's now consider the LTE numerology \cite{LTEBook}. For a carrier frequency of
$2$GHz, as the user's moving speed is around $30$km/h, we see the max Doppler
shift is about $f_D=55$Hz. Considering each OFDM symbol lasts $T_s=66.67\mu$s, the
channel sampling frequency can be chosen as $f_s=1/(3T_s)=5$kHz for OFDM. The
maximum normalized Doppler shift is then $F_{k}=f_D/f_s=0.011\ll1$. Thus, when the
receiving signal-to-noise ratio (SNR) of the pilots is not too small, e.g. $>-10$dB,
we only need to consider the first case in (\ref{mse c}). The following corollary
can be established from Proposition 3 (see Appendix B for detailed proof):\\
\noindent{\bf Corollary 3.1}: {\it In a communication system with $\pi F_k\ll\rho_k/\sigma^2$, under the classical Clarke's fading, as the orthogonality
conditions in (\ref{ortho}) are satisfied and the length of the channel
observations: $P$ goes large,
under the assumptions AS1$\sim$3, the MSE of each element of ${\bm h}_{k,m}$ can be
approximated as follows:
\begin{equation}\label{mse appro}
  \lim_{P\rightarrow\infty}{\sf MSE}_k=\frac{2 F_k}{\rho_k/\sigma^2},
\end{equation}
and the following processing gain\footnote{The processing gain here
refers to the amount of SNR improvement during the channel estimation relative to
the observation SNR \cite{DigitalComBook}. } in dB scale can be achieved:
\begin{equation}\label{PG_appro}
  \lim_{P\rightarrow\infty}G_k=10\log_{10}\left(\frac{1}{2F_k}-\frac{1}{\rho_k/\sigma^2}\right).
\end{equation} }

\vspace{-0.5cm}
\section{Simulated Performance}\label{secPerformance}

\begin{figure}[t]
\centering
\epsfig{file=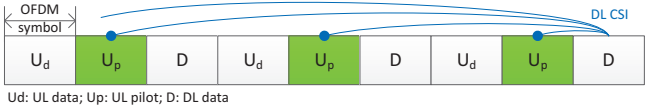,width=0.48\textwidth}
\caption{Simulated TDD configuration.}
\label{TDDconfig}
\vspace{-0.5cm}
\end{figure}

\begin{figure}[ht!]
\centering
\epsfig{file=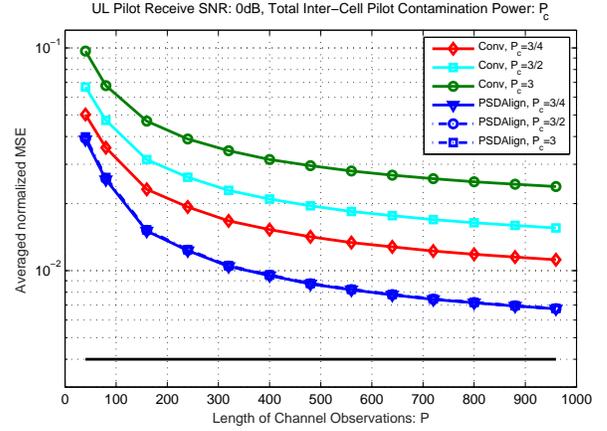,width=0.43\textwidth}
\caption{Average nMSE of the $8$ users' UL channel estimates.
(Conv: $8\times8$ Hadamard pilot sequences;
PSDAlign: Proposed PSD aligning pilots with $\tau_k/P=3/8+k/36$, $k=1,...,8$;
Analytical: Result from Corollary 3.1. Each user's pilot SNR is $0$dB at each BS
receive antenna.)}
\label{MMSE}
\vspace{-0.2cm}
\end{figure}

\begin{figure}[ht!]
\centering
\epsfig{file=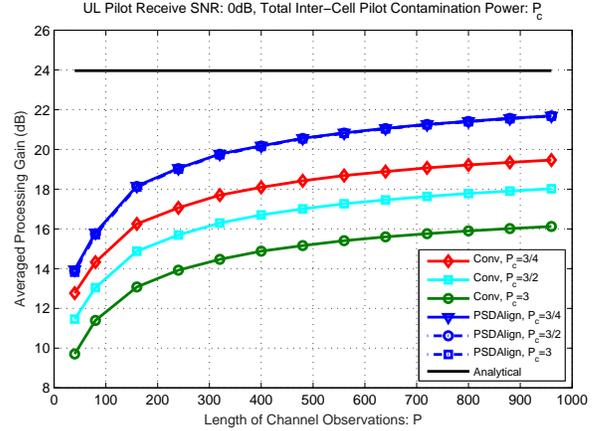,width=0.43\textwidth}
\caption{Achieved average processing gains.}
\label{processingGain}
\vspace{-0.2cm}
\end{figure}

\begin{figure}[ht!]
\centering
\epsfig{file=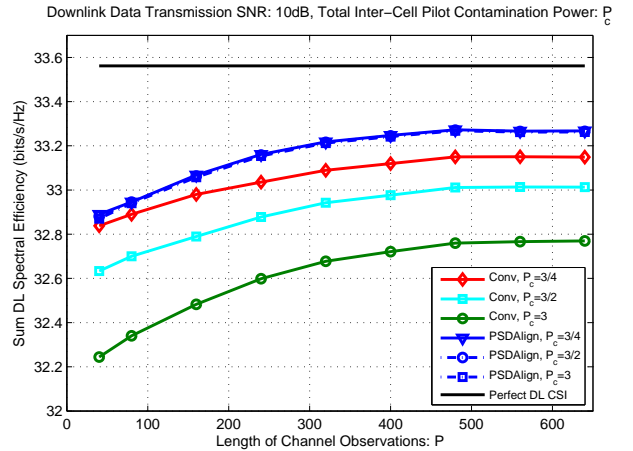,width=0.45\textwidth}
\caption{Achieved DL sum spectral efficiency.}
\label{cap comp}
\end{figure}

To verify our orthogonal designs for Doppler, we simulate a TDD reciprocal massive
MIMO system as configured in Fig. \ref{TDDconfig}, where $8$ users experience the
Jakes's flat-fading channels with the same Doppler frequency $f_D=10$Hz and the channel gains towards the serving BS.
Inter-cell pilot contamination is modelled as a stationary random process with
uniform PSD in the discrete frequency range: $[-3/8, +3/8]$.
The following system parameters are assumed during the simulations:
\begin{itemize}
  \item OFDM symbol duration\footnote{Since all the simulations here are on a particular subcarrier, the exact number of subcarriers within one OFDM symbol does not really matter.}
  : $T_s= 66.67\mu$s; channel sampling frequency:
  $f_s= \frac{1}{3T_s}= 5$kHz;
  \item Antenna array size at the BS: $M=128$; different antennas at the BS are
  assumed independent.
\end{itemize}
Fig. \ref{MMSE} compares the channel estimation MSE normalized by the average
channel power (nMSE) with different pilot sequences designs.
Fig. \ref{processingGain} depicts the achieved processing gains
of our scheme and the conventional ones. Clearly, our proposed pilot designs exhibit
significant improvement in both MSE and processing gain with respect to the
convention pilots. In Fig. \ref{cap comp}, we plot the sum downlink (DL) spectral
efficiency when the BS performs the matched-filter beamforming \cite{marzetta10twc} to the
served users. Exploiting the TDD reciprocity, the DL channel states are obtained from the estimated UL channels with the previous UL pilots as illustrated in
Fig. \ref{TDDconfig}.
We see more accurate DL channel state information (CSI) at the BS enabled by
aligning the PSDs
appropriately translates to higher DL spectral efficiency in the presence of
time-varying pilot contamination.

\section{Conclusion}\label{conclusion}
In this paper, we have developed a novel pilot design principle in the case of
Doppler fading. Through flexibly aligning the PSDs, pilot decontamination can be
achieved and more users' channels can be sounded simultaneously even with high
Doppler. Meanwhile, we have derived analytical formulae characterizing the channel
estimation MSE performance with our proposed pilot designs. Numerical simulations
corroborate our designs and demonstrate that our proposed designs outperform the
conventional designs significantly as the length of the channel observations
$P$ goes large. In practice, to obtain the autocorrelation matrix at the BS involves
an overhead proportional to $P^2$, which would limit the maximum length of the
channel observations.

The proposed PSD aligning can serve as a new complementary design
philosophy for the UL pilots to address the notorious pilot contamination in massive
MIMO. In particular, our proposed PSD aligning can be naturally combined with those
decontamination schemes exploiting the spatial separability to handle the cases with
overlapping AoAs.

\section*{Appendix A: Proof of Proposition 3}
When $P$ goes to infinity, by approximating ${\bm R}_k$ (${\bm R}_g$) with a
circulant matrix ${\bm C}_k$ (${\bm C}_g$), we can rewrite (\ref{err int}) as
\begin{equation}
  \begin{aligned}
  &{\sf E}[{\bm\epsilon}{\bm\epsilon}^{\dag}]={\bm R}_k-\rho_k{\bm R}_k\cdot\\
  &\left(\sigma^2{\bm I}+\rho_k{\bm R}_k+\sum_{g=1,g\ne k}^{K}\rho_g{\bm P}_{k,g}{\bm R}_g{\bm P}_{k,g}^{\dag}\right)^{-1}{\bm R}_k\\
  &={\bm F}^{\dag}\Bigg({\bm\Lambda}_k-\rho_k{\bm\Lambda}_k\cdot\\
  &\Big(\sigma^2{\bm I}+\rho_k{\bm\Lambda}_k+\sum_{g=1,g\ne k}^{K}\rho_g{\bm\Theta}_{k,g}{\bm\Lambda}_g{\bm\Theta}_{k,g}^{\dag}\Big)^{-1}{\bm\Lambda}_k\Bigg){\bm F}\\
  &={\bm F}^{\dag}\Bigg({\bm\Lambda}_k-\rho_k{\bm\Lambda}_k\cdot\\
  &\Big(\sigma^2{\bm I}+\rho_k{\bm\Lambda}_k+\sum_{g=1,g\ne k}^{K}\rho_g\tilde{\bm\Lambda}_g\Big)^{-1}{\bm\Lambda}_k\Bigg){\bm F}.
  \end{aligned}
\end{equation}
Following the results in \cite{adhikary13tit}, the MSE of each element of ${\bm h}_{k,m}$ can then be approximated as:
\begin{equation}
\begin{aligned}
  &\lim_{P\rightarrow\infty}{\sf MSE}_k=\lim_{P\rightarrow\infty}\frac{1}{P}{\tt Tr}({\sf E}[{\bm\epsilon}{\bm\epsilon}^{\dag}])\\
  &=\lim_{P\rightarrow\infty}\frac{1}{P}{\tt Tr}\Bigg({\bm F}^{\dag}\bigg({\bm\Lambda}_k-\rho_k{\bm\Lambda}_k\cdot\\
  &\Big(\sigma^2{\bm I}+\rho_k{\bm\Lambda}_k+\sum_{g=1,g\ne k}^{K}\rho_g\tilde{\bm\Lambda}_g\Big)^{-1}{\bm\Lambda}_k\bigg){\bm F}\Bigg)\\
  &=\lim_{P\rightarrow\infty}\Bigg(1-\sum_{p=1}^{P}\frac{\lambda_{k,p}^2\rho_k}{\lambda_{k,p}\rho_k+\sum\limits_{g=1,g\ne k}^{K}\tilde{\lambda}_{g,p}\rho_g+\sigma^2}\Bigg)\\
  &=1-\int_{-\frac{1}{2}}^{\frac{1}{2}}\frac{S_k^2(\xi)\rho_k}{S_k(\xi)\rho_k+\sum_{g=1,g\ne k}^{K}\tilde{S}_{g}(\xi)\rho_g+\sigma^2} d\xi,
\end{aligned}
\end{equation}
where $\lambda_{k,p}$ is the $p$-th eigenvalue of ${\bm \Lambda}_k$, $S_k(\xi)$ denotes the PSD of user-$k$, and $\tilde{S}_g(\xi):=S_{g}(\xi-(\tau_g-\tau_k)/{P})$.

\section*{Appendix B: Derivation of Corollary 3.1}
In a communication system with $\pi F_k\ll\rho_k/\sigma^2$, we can focus on
the case with $\alpha\ll1$ in (\ref{mse c}). Carrying out the Taylor series
expansion in the case with $\alpha\ll1$, we can obtain
\begin{equation}\label{MSE_Taylor}
\begin{aligned}
  &\lim_{P\rightarrow\infty}{\sf MSE}_k=
  \frac{2}{\pi}\alpha-\frac{1}{2}\alpha^2+\frac{4}{3\pi}\alpha^3+o(\alpha^3).
\end{aligned}
\end{equation}
Since $\alpha\ll1$, the first-order term in (\ref{MSE_Taylor}) already 
provides a close approximate. Thus we have the desired result as follows:
\begin{equation}
\begin{aligned}
  \lim_{P\rightarrow\infty}{\sf MSE}_k
  \approx \frac{2}{\pi}\alpha
  =\frac{2F_k}{\rho_k/\sigma^2}.
\end{aligned}
\end{equation}

%\newpage
\bibliographystyle{IEEE}

\end{document}